\documentclass[a4paper, 11pt, oneside]{article} 

\usepackage{geometry}
\usepackage{authblk}
\usepackage[utf8]{inputenc} 
\usepackage[T1]{fontenc} 
\usepackage{fouriernc} 
\usepackage{epsfig}
\usepackage{listings}
\usepackage[normalem]{ulem}
\usepackage{algorithmic}
\usepackage{algorithm}
\usepackage{makecell}
\usepackage{epsfig}
\usepackage{listings}
\usepackage[normalem]{ulem}
\usepackage{algorithmic}
\usepackage{algorithm}
\usepackage{makecell}
\usepackage{fancyhdr}
\usepackage{hyperref}
\usepackage{amsmath}
\usepackage{graphicx}
\usepackage{amsmath}
\usepackage{fancyhdr}
\usepackage{hyperref}
\usepackage{amsmath}
\usepackage{graphicx}
\usepackage[normalem]{ulem}
\usepackage{algorithmic}
\usepackage{algorithm}
\usepackage{multirow}
\usepackage{epsfig}
\usepackage{listings}
\usepackage[normalem]{ulem}
\usepackage{algorithmic}
\usepackage{algorithm}
\usepackage{makecell}
\usepackage{fancyhdr}
\usepackage{hyperref}
\usepackage{amsmath}
\usepackage{color}
\usepackage{cite}
\usepackage{amsmath,amssymb,amsfonts}
\usepackage{algorithmic}
\usepackage{graphicx}
\usepackage{textcomp}
\usepackage{xcolor}
\newenvironment{sequation}{\begin{equation}\small}{\end{equation}}

\begin{document}

\begin{titlepage} 

	\centering 
	
	\scshape 
	
	\vspace*{\baselineskip} 
	
	
	\rule{\textwidth}{1.6pt}\vspace*{-\baselineskip}\vspace*{2pt} 
	\rule{\textwidth}{0.4pt} 
	
	\vspace{0.75\baselineskip} 
	
	{\LARGE BOPS, Not FLOPS! \\A New Metric and Roofline Performance Model For Datacenter Computing\\} 
	
	\vspace{0.75\baselineskip} 
	
	\rule{\textwidth}{0.4pt}\vspace*{-\baselineskip}\vspace{3.2pt} 
	\rule{\textwidth}{1.6pt} 
	
	\vspace{2\baselineskip} 
	
	
	
	\vspace*{3\baselineskip} 
	
	
	Edited By
	
	\vspace{0.5\baselineskip} 
	
	{\scshape\Large Lei Wang\\Jianfeng Zhan \\ Wanling Gao\\ KaiYong Yang\\ ZiHan Jiang \\ Rui Ren \\Xiwen He \\Chunjie Luo} 
	
	\vspace{0.5\baselineskip} 

	\vfill 
	
	
	\epsfig{file=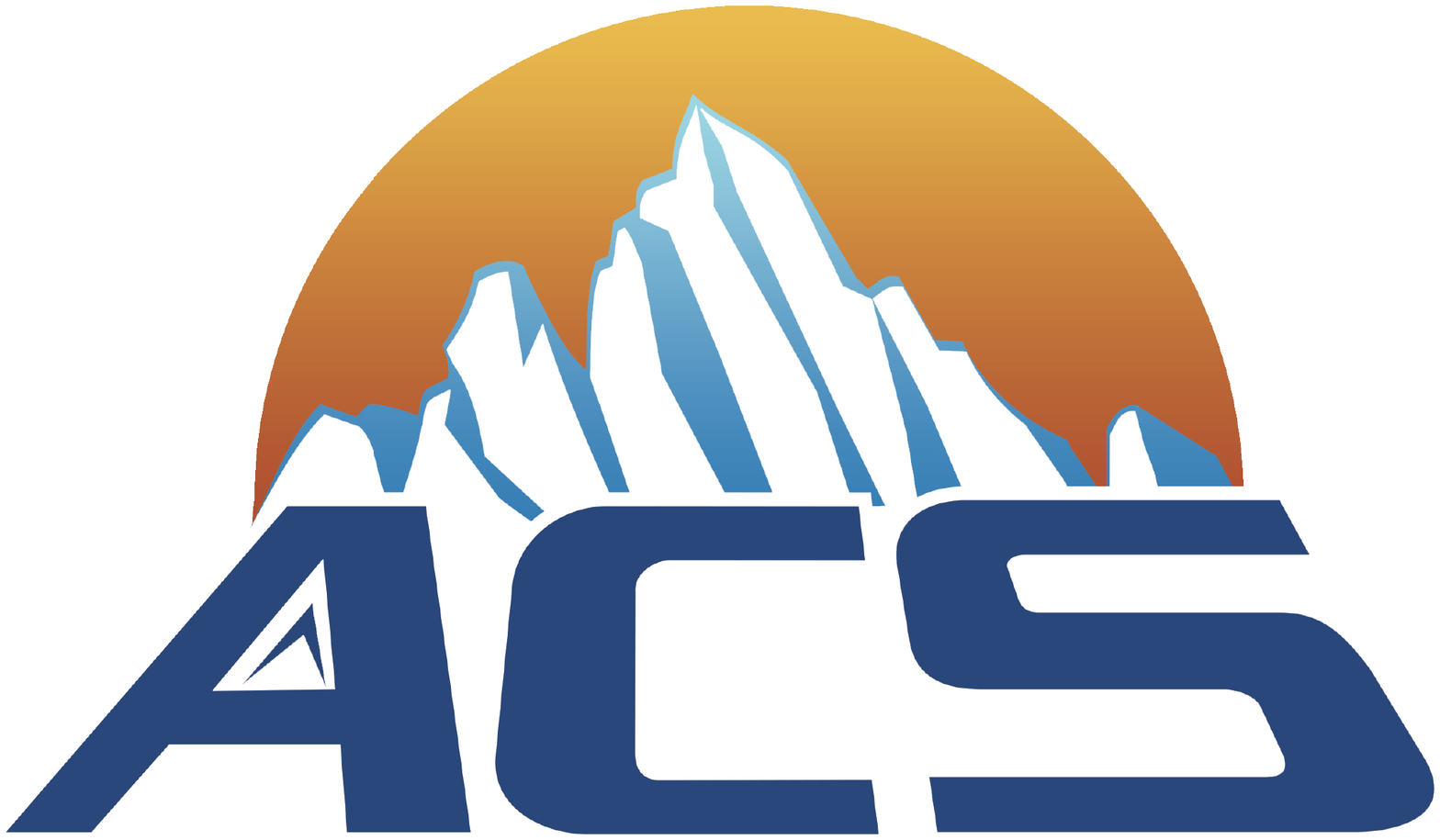,height=5cm,angle=270}
	\textit{\\Software Systems Laboratory (SSL), ACS\\ICT, Chinese Academy of Sciences\\Beijing, China} 
	\vspace{5\baselineskip} 

	
	{\large Aug, 15, 2019} 

\end{titlepage}


\title{BOPS, Not FLOPS! A New Metric and Roofline Performance Model For Datacenter Computing}

\author{Lei Wang, Jianfeng Zhan, Wanling Gao, KaiYong Yang, ZiHan Jiang, Rui Ren, Xiwen He, Chunjie Luo}

\maketitle
\begin{abstract}
For emerging datacenter (in short, DC) workloads, such as online Internet services or offline data analytics, how to evaluate the upper bound performance and provide apple-to-apple comparisons are fundamental problems. To this end, an unified computation-centric metric is an essential requirement. FLOPS (FLoating-point Operations Per Second) as the most important computation-centric performance metric, has guided computing systems evolutions for many years. However, our observations demonstrate that the average FLOPS efficiency of the DC workloads is only 0.1\%, which implies that FLOPS is inappropriate for DC computing. To address the above issue, inspired by FLOPS, we propose BOPS (Basic Operations Per Second), which is the average number of BOPs (Basic OPerations) completed per second, as a new computation-centered metric. We conduct the comprehensive analysis on the characteristics of seventeen typical DC workloads and extract the minimum representative computation operations set, which is composed of integer and floating point computation operations of arithmetic, comparing and array addressing. Then, we propose the formalized BOPS definition and the BOPS based upper bound performance model. Finally, the BOPS measuring tool is also implemented. To validate the BOPS metric, we perform experiments with seventeen DC workloads on three typical Intel processors platforms. First, BOPS can reflect the performance gap of different computing systems, the bias between the peak BOPS performance (obtaining from micro-architecture) gap and the average DC workloads' wall clock time gap is no more than 10\%. Second, BOPS can not only perform the apple-to-apple comparison, but also reflect the upper bound performance of the system. For examples, we analyze the BOPS efficiency of the Redis (the online service) workload and the Sort (the offline analytics) workload. And using the BOPS measuring tool--Sort can achieve 32\% BOPS efficiency on the experimental platform. At last, we present two use cases of BOPS. One is the BOPS based system evaluation, we illustrate that using BOPS, we can compare performance of workloads from multiple domains, which include but are not limited to Big data, AI, HPC, OLTP and OLAP, then we can determine which kind of workload is more suitable for the such DC system. The other is the BOPS based DC workload optimizations, we show that under the guiding of the BOPS based upper bound performance model, named \emph{DC-Roofline}, the Sort workload achieves 4.4X performance improvements. Furthermore, we propose an optimization methodology for the real-world DC workloads, which is based on the DC-Roofline model. Under the guidance of the proposed methodology, we optimize Redis (a typical real-world DC workload) by 1.2X.

\end{abstract}
\section{Introduction}
To perform data analysis or provide Internet services, more and more organizations are building internal datacenters, or renting hosted datacenters. As a result, DC (datacenter) computing has become a new paradigm of computing. The proportion of DC has outweighed HPC (High Performance Computing) in terms of market share (HPC only takes 20\% of total)~\cite{DCGrowth}.  How to evaluate the performance efficiency and provide apple-to-apple comparisons are fundamental problems. There is an urgent need for a unified metric. For HPC, FLOPS is a powerful metric and has promoted its rapid evolution and optimization over a period of decades~\cite{dongarra2003Linpack}. However, for DC, there is still no such metric.

Generally, the wall clock time is used as a ground truth metric for the computer system. Based on it, the performance metrics are classified into two categories. One is the user-perceived metric, which can be intuitively perceived by the user, such as requests per second~\cite{ycsb}, sorting number per second ~\cite{sortben}. The other is the computation-centered metrics, which are related to specific computation operations, such as FLOPS (FLoating-point Operations Per Second).

User-perceived metrics can be intuitively perceived by the user. But, user-perceived metrics have two limitations. First, user-perceived metrics are hard to measure the upper bound performance of computer systems, which is the foundation of the quantitative evaluation. For example, for the matrix multiply workload, the deep optimized version gains 62,000X out performance of the original Python version on the same Intel multi-core processor platform~\cite{hennessy2019new}. So, we wonder the performance efficiency and the upper bound performance of the matrix multiply workload on this platform. Second, different user-perceived metrics cannot be used to perform the apple-to-apple comparison. For example, requests per second and sorting number per second cannot be used for comparison. We cannot obtain the performance efficiency for different type of workloads on the target system.

Computation-centric metrics solve the above limitations. Different workloads can perform the apple-to-apple comparisons. Furthermore, the performance numbers of the metric can be measured by the micro-architecture of the system, the specific micro benchmark and the real-world workload. By using these different numbers, we can build the upper bound model, which allows us to understand the upper bound performance of the computer system. For example, FLOPS motivates continuously exploring to achieve upper bound performance. Also, the winner of Gordon Bell prize and TOP500 ranking represents the best FLOPS performance currently ~\cite{top50020151120}. However, FLOPS is insufficient for DC anymore. Our experiments show that the average FLOPS efficiency is only 0.1\% for DC workloads, so that it cannot represent the actual execution performance for DC. OPS (operations per second) is another computation-centric metric. OPS~\cite{nakajima200640gops} is initially proposed for digital processing systems. The definitions of OPS are extended to the artificial intelligence processor~\cite{abadi2016tensorflow:,Jouppi2017In,Liu2016Cambricon,chen2014dadiannao:}. All of them are defined in terms of one or more fixed operations, such as the specific matrix addition operation. However, these operations are only a fraction of diverse operations in DC workloads.

In this paper, inspired by FLOPS~\cite{dongarra2003Linpack} and OPS~\cite{nakajima200640gops}, \underline{B}asic \underline{OP}erations per \underline{S}econd (BOPS for short) is proposed to evaluate DC computing systems. The contributions of the paper are described as follows.

First, Based on workload characterizations of seventeen typical DC workloads, we find that DC workloads are data movement dominated workloads, which have more integer and branch operations. Then, following the rule of choosing a representative minimum operation subset, we define BOPs as the integer and floating point computation operations of arithmetic, comparing and array addressing (related with data movement). For the quantitative evaluation, the formalized BOPS definition and the BOPS based upper bound performance model are also given, Finally, we implement the Sort workload as the first BOPS measuring tool.

Second, we validate the BOPS metric on three typical Intel processors systems with seventeen typical DC workloads. Results show that BOPS can reflect the performance gap of different systems, and the bias between the peak BOPS performance (obtaining from micro-architecture) gap and the average DC workloads' wall clock time gap is no more than 10\%. Furthermore, BOPS can not only perform the apple-to-apple comparison, but also reflect the upper bound performance of the system. The Redis (the online service workload) and the Sort (the offline analytics workload) can perform the BOPS performance comparison. The BOPS efficiency of the Sort workload achieves 32\% of the peak performance of the Intel Xeon E5645 platform, and the attained upper bound performance efficiency (calculating by the BOPS based upper bound model) of Sort achieves 68\%.

Third, we illustrate the use cases of BOPS. First, BOPS based upper bound model, named \emph{DC-Roofline}, can help to guide the optimization of the
DC workloads. For the Sort workload, the performance improvement achieves 4.4X. Second, the BOPS metric is also suitable for typical HPC benchmarks, which include HPL, Graph500 and Stream workloads. We illustrate that using BOPS to evaluate the computer system from multiple application domains, which include but are not limited to Big data, AI, HPC, OLTP and OLAP.

Forth, a real-world DC workload always has million lines of codes and tens of thousands of functions, so it is not easy to use the DC-Roofline model directly. We propose a new optimization methodology. We profile the hotspot functions of the real-world workload and extract the corresponding kernel workloads. The real-world application gains performance benefit from merging optimizations methods of kernel workloads, which are under the guidance of DC-Roofline. Through experiments, we demonstrate that Redis---a typical real-world workload gains performance improvement by 1.2X.




The remainder of the paper is organized as follows. Section 2 summarizes the related work. Section 3 states background and motivations. Section 4 defines BOPS, and reports how to use it. Section 5 is the evaluations of BOPS. Section 6 is the use case of BOPS. Section 7 draws a conclusion.
\section{Related Work}
The performance metrics can be classified into two categories. One is the user-perceived metric, another is the computation-centric metric.

User-perceived metrics can be further classified into two categories: one is the metric for the whole system, and the other is the metric for components of the system. The examples of the former include data sorted in one minute (MinuteSort), which measures the sorting capability of a system~\cite{sort}, and transactions per minute (TPM) for the online transaction system~\cite{tpcc}. The examples of the latter include the SPECspeed/SPECrate for the CPU component~\cite{speccpu}, the input/output operations per second (IOPS) for the storage component~\cite{josephson2010dfs}, and the data transfer latency for the network component~\cite{cardwell2000modeling}.

There are many computation-centric metrics. FLOPS (FLoating-point Operations Per Second) is a computation-centric metric to measure the computer system, especially in field of the scientific computing that makes heavy use of floating-point
calculations~\cite{dongarra2003Linpack}. The wide recognition of FLOPS indicates the maturation of high performance computing. MIPS (Million Instructions Per Second)~\cite{jain1991art} is another famous computation-centric metric, which is defined as the million number of instructions the processor can process in a second. The main limitation of MIPS is that it is
architecture-dependent. There are many derivatives of the MIPS, including MWIPS and DMIPS~\cite{pesovic2012benchmarking}, which use synthetic workloads to evaluate the floating point operations and integer
operations, respectively. The WSMeter metric\cite{lee2018wsmeter}, which is defined as the quota-weighted sum of MIPS of a job, is also a derivative of MIPS, and hence it is also architecture-dependent. Unfortunately, modern DCs are heterogeneous, which consist of different types of hardware. OPS (Operations Per Second) is another computation-centric metric. OPS~\cite{nakajima200640gops} is initially proposed for digital processing systems, which is defined as the 16-bit addition operations per second. The definitions of OPS are then extended to Intel Ubiquitous High Performance Computing~\cite{carter2013runnemede} and artificial intelligence processors, such as Tensor Processing Unit~\cite{abadi2016tensorflow:,Jouppi2017In} and Cambricon processor~\cite{Liu2016Cambricon,chen2014dadiannao:}. All of these definitions are in terms of one or more fixed operations. For example, Operations are 8-bit matrix multiplication operations in TPU and 16-bit integer operations in Cambricon processor, respectively. However, the workloads of modern DCs are comprehensive and complex, and the bias to one or more fixed operations can not ensure the evaluation fairness.

For each kind of metrics, the corresponding tools or benchmarks~\cite{hennessy2012computer} are proposed to calculate the values. For user-perceived metrics---SPECspeed/SPECrate, SPECCPU is the benchmark suite~\cite{speccpu} to measure the CPU component. For computation-centric metrics, Whetstone~\cite{harbaugh1984timing} and Dhrystone~\cite{weicker1984dhrystone:} are the measurement tools for MWIPS and DMIPS, respectively. HPL~\cite{dongarra2003Linpack} is a widely used measurement tool for FLOPS.

For computation-centric metrics, the Roofline model~\cite{williams2009roofline} is the famous performance model. The Roofline model can depict the upper bound performance of given workloads, when different optimization strategies are adopted to the target system. The original Roofline model~\cite{williams2009roofline} adopts FLOPS as the performance metric.
\section{Background and Motivations}
\subsection{Background}
\subsubsection{The Computation-centric Metric}\label{bops_measure}
The computation-centric metric is the key element to quantitatively depict the performance of the system ~\cite{hennessy2012computer}. The computation-centric metric can be calculated at the source code level of applications,
which is independent with the underlying system implementations. Also, it can be calculated at the binary code
level of softwares and the instruction level of hardware, respectively. So it is effective for the co-design across different layers. Generally, a computation-centric metric has performance upper bound on the specific architecture according to the micro-architecture design. For example, the peak FLOPS is computed as follows.
\begin{sequation}
FLOPS_{Peak}=Num_{CPU}*Num_{Core}*Frequency*Num_{Floatingpoint Operations Per Cycle}
\end{sequation}
For example, in our experimental platform, the Intel Xeon E5645 is Westmere architecture, which is equipped with four-issue and out-of-order pipeline. In the E5645 platform, the CPU number is 1, the Core number is 6, the frequency of each CPU Core is 2.4GHZ, the floating point operations per cycle of each CPU Core is 4 (four-issue and two independent 128-bit SSE FPUs). So, the E5645's peak FLOPS is 57.6 GFLOPS.

The measurement tool is used to measure the performance of systems and architectures in terms of metric values, and report the gap between the real value and the theoretical peak one.
For example, HPL~\cite{dongarra2003Linpack} is a widely used measurement tool in terms of FLOPS.
The FLOPS efficiency of a specific system is the ratio of the HPL's FLOPS to the peak FLOPS.
\begin{sequation}
FLOPS_{Efficiency}=FLOPS_{Real}/FLOPS_{Peak}
\end{sequation}
In our experiments, the real FLOPS obtaining from the HPL benchmark is 38.9 GFLOPS, and the FLOPS efficiency of the E5645 platform is 68\%.
\subsubsection{The Upper Bound Performance Model}
The bound and bottleneck analysis can be built under the computation-centric metric. For example, the Roofline model ~\cite{williams2009roofline} is a famous upper bound model based on FLOPS. There are many system optimization works ~\cite{kamil2010an} ~\cite{williams2012optimization}, which are performed based on the Roofline model in the HPC domain.
\begin{sequation}
FLOPS_{AttainedPeak}=min(OI*MemBand_{Peak},FLOPS_{Peak})
\end{sequation}
The above equation of the Roofline model indicates that the attained workload performance bound of a specific platform is limited by the computing capacity of the processor and the bandwidth of the memory. $\textstyle{FLOPS_{Peak}}$ and $\textstyle{MemBand_{Peak}}$ are the peak performance of the platform, and the operation intensity (i.e., $\textstyle{OI}$) is the total number of floating point operations divided by the total byte number of memory access. If the attained peak FLOPS is $\textstyle{OI*MemBand_{Peak}}$, the bottleneck is the bandwidth of the memory. Otherwise, the bottleneck is the computing capacity. For example, the OI of the HPL benchmark is 6.1, the peak memory bandwidth is 13.8GB/s, the  $\textstyle{FLOPS_{Peak}}$ is 57.6 GFLOPS. So, the attained peak FLOPS of the HPL is 57.6 GFLOPS, the bottleneck is the computing capacity and the attained FLOPS efficiency ($\textstyle{FLOPS_{Real}/FLOPS_{AttainedPeak}}$) is 68\% too. Furthermore, based on the Equation 3, to identify the bottleneck and guide the optimization, the ceilings (for example, the ILP and SIMD optimizations) can be added to provide the performance tuning guidance~\cite{williams2009roofline}.
\subsubsection{DCMIX}
We choose the DCMIX~\cite{xiong2018dcmix} as benchmarks for DC computer systems. DCMIX is designed for modern DC computing systems, which has 17 typical DC workloads (including online service and data analysis workloads). Latencies of DCMIX workloads are ranged from microseconds to minutes. The applications of DCMIX involve Big Data, artificial intelligence (AI), OLAP, and OLTP. As shown in Table~\ref{WorkloadsEx1}, there are two categories of benchmarks in the DCMIX, which are Micro-Benchmarks (kernel workloads) and Component benchmarks (real DC workloads).
\begin{table}[H]
\center
\begin{footnotesize}
\caption{Workloads of the DCMIX}
\label{WorkloadsEx1}
\begin{tabular}{|l|l|l|l|}
        \hline
        \textbf{Name} & \textbf{Type} & \textbf{Domain} & \textbf{Category} \\
        \hline
        Sort & offline analytics & Big Data& MicroBench \\
        \hline
        Count& offline analytics & Big Data& MicroBench \\
        \hline
        MD5 & offline analytics &  Big Data & MicroBench\\
        \hline
        MatrixMultiply & offline analytics & AI& MicroBench \\
        \hline
        FFT & offline analytics &AI & MicroBench \\
        \hline
        Union & offline analytics & OLAP &  MicroBench \\
        \hline
        Redis & online service & OLTP & Component  \\
        \hline
        Xapian & online service & Big Data &  Component  \\
        \hline
        Masstree & online service & Big Data & Component  \\
        \hline
        Bayes & offline analytics & Big Data& Component \\
        \hline
        Img-dnn & online service & AI & Component  \\
        \hline
        Moses & online service & AI & Component  \\
        \hline
        Sphinx& online service & AI & Component  \\
        \hline
        Alexnet & offline analytics & AI& Component\\
        \hline
        Convolution & offline analytics & AI& Component\\
        \hline
        Silo & online service & OLTP & Component  \\
        \hline
        Shore & online service & OLAP & Component  \\
        \hline
\end{tabular}
\end{footnotesize}
\end{table}

\subsection{Motivations}
\subsubsection{Requirements of the DC computing metric}
We define the requirements from the following perspectives. First, \textbf{the metric should reflect the performance gaps of different DC systems}. The wall clock time metric always reflect the performance gaps of different systems. Also, the computation-centric metric should preserve this characteristic. We can use the bias between the computing metric gap and the wall clock time gap to evaluate this requirement. Second, \textbf{the metric should reflect the upper bound performance of the DC system and facilitate measurements}. Focusing on different system design, the metric should be sensitive to design decisions and reflect theoretical performance upper bound. Then, the gap between real and theoretical values is useful to understand the performance bottlenecks and guide the optimizations.

\subsubsection{Experimental Platforms and workloads}
We choose DCMIX as DC workloads. Three systems equipped with three different Intel processors are chosen as the experimental platforms, which are Intel Xeon E5310, Intel Xeon E5645 and Intel Atom D510. The two former processors are typical brawny-core processors (OoO execution, four-wide instruction issue), while Intel Atom D510 is a typical wimpy-core processor (in-order execution, two-wide instruction issue). Each experimental platform is equipped with one node. The detailed settings of platforms are shown in Table~\ref{hwconfigeration1}.
\begin{table}[ht]
\center
\caption{Configurations of Hardware Platforms.}
\begin{footnotesize}
\begin{tabular}{|c|c|c|c|}
  \hline
  \multicolumn{2}{|c|}{CPU Type} & \multicolumn{2}{|c|}{CPU Core} \\ \hline
  \multicolumn{2}{|c|}{Intel \textregistered Xeon E5645}  &\multicolumn{2}{|c|}{6 cores@2.4 GHZ} \\ \hline
  L1 DCache &L1 ICache &L2 Cache &L3 Cache \\ \hline
6 $\times$ 32 KB& 6 $\times$ 32 KB&6 $\times$ 256 KB& 12 MB \\ \hline
  \hline
  \multicolumn{2}{|c|}{CPU Type} & \multicolumn{2}{|c|}{CPU Core} \\ \hline
  \multicolumn{2}{|c|}{Intel \textregistered Xeon E5310}  &\multicolumn{2}{|c|}{4 cores@1.6 GHZ} \\ \hline
L1 DCache &L1 ICache &L2 Cache &L3 Cache \\ \hline
4 $\times$ 32 KB& 4 $\times$ 32 KB&2 $\times$ 4 MB& None \\ \hline
  \hline
  \multicolumn{2}{|c|}{CPU Type} & \multicolumn{2}{|c|}{CPU Core} \\ \hline
  \multicolumn{2}{|c|}{Intel \textregistered Atom D510}  &\multicolumn{2}{|c|}{2 cores@1.6 GHZ} \\ \hline
L1 DCache &L1 ICache &L2 Cache &L3 Cache \\ \hline
2 $\times$ 24 KB& 2 $\times$ 32 KB&2 $\times$ 512 KB&None  \\ \hline
\end{tabular}
\end{footnotesize}
\label{hwconfigeration1}
\end{table}

\subsubsection{The Limitation of FLOPS for DC}
Corresponding with requirements of the DC computing metric, we evaluate the FLOPS from two aspects. One is reflecting the performance gaps of different DC systems, another is reflecting the upper bound performance.

The performance gaps are from three folds. First, the performance gaps between E5310 and E5645, the peak FLOPS performance gap is 2.3X (25.6 GFLOPS v.s. 57.6 GFLOPS), and the gap of the average wall clock time is 2.1X. The bias is 9\%. Second, the performance gaps between D510 and E5645, the peak FLOPS gap is 12X (4.8 GFLOPS v.s. 57.6 GFLOPS), and the gap of the average wall clock time is 7.4X. The bias is 62\%. Third, for the performance gaps between D510 and E5310, the peak FLOPS gap is 5.3X, the gap of the average user-perceived performance metrics is 3.4X. The bias is 60\%. \textbf{The bias of the peak FLOPS performance gap and the average wall clock time gap between the two systems equipped with Intel Xeon or Intel Atom processors is more than 60\%}. This is because that E5645\&E5310 and D510 are totally different micro-architecture platforms, E5645\&E5310 are designed for high performance floating point computing, while D510 is a low power microprocessor for mobile computing. But, DC workloads are data movement intensive workloads, so the performance gaps between Xeon and Atom become narrowed.

For reflecting the upper bound performance, we use six microbenchmarks of DCMIX to reveal the limitations of FLOPS for DC. The details are shown in Table ~\ref{WorkloadsEx1456}. The FLOPS of DC workloads is only 0.08 GFLOPS on average (only 0.1\% of the peak). We use the Roofline model (Equation 3) to measure the attained upper bound performance and locate the bottlenecks of DC workloads. In our experiment platform--Intel E5645, the peak FLOPS is 57.6 GFLOPS, and the peak memory bandwidth (before optimizations) is 13.2GB/s. s. From Table ~\ref{WorkloadsEx1456}, we can observe that, first, the operation
intensity (OI) for six workloads is very low, and the average number is only 0.2. Second, the Roofline model (Equation
3) indicates that the bottleneck of all six workloads is the bandwidth of the memory. Furthermore, under the hint of the
Roofline model, we increase the memory bandwidth through hardware pre-fetching, and the peak memory bandwidth increases
from 13.2GB/s to 13.8GB/s. However, only Union and Count gain obvious performance improvement, which are 16\% and 10\%, respectively. For the other four workloads, the average performance improvement is no more than 3\%, which indicates the bandwidth of memory is not the real bottleneck. So, FLOPS is not suitable to obtain the upper bound performance of workloads on the target system.

\begin{table}[H]
\center
\begin{footnotesize}
\caption{DC Workloads under The Roofline Model.}
\label{WorkloadsEx1456}
\begin{tabular}{|l|l|l|l|}
\hline
        Workload   & FLOPS&OI & Bottleneck         \\
  \hline
      Sort   & 0.01G &0.01 &Memory Access       \\
  \hline
      Count   & 0.01G & 0.01 &Memory Access       \\
  \hline
      MD5   & 0.01G &0.02 &Memory Access       \\
  \hline
      MatrixMultiply   & 0.2G &0.6 &Memory Access       \\
  \hline
      FFT   & 0.2G &0.5  &Memory Access       \\
  \hline
      Union   & 0.05G &0.1  &Memory Access       \\
  \hline
  \end{tabular}
\end{footnotesize}
\end{table}

\subsubsection{The characteristics of DC Workloads}
In order to define the new metric for the DC computing, we perform a careful workload characterization of DC workloads firstly. We choose the DCMIX as the DC workloads. For traditional benchmarks, we choose HPCC, PARSEC, and SPECPU. We have used HPCC 1.4, which is a representative HPC benchmark suite, for the experiment. We run all of the seven benchmarks in HPCC. PARSEC is a benchmark suite composed of multi-threaded programs, and we deploy PARSEC 3.0. For SPEC CPU2006, we run the official floating-point benchmark (SPECFP) applications with the first reference inputs. The experimental platform is the Intel Xeon E5645.

We choose GIPS (Giga-Instructions per Second) and GFLOPS (Giga-Floating point Operations Per Second) as the performance metrics. Corresponding to performance metrics, we choose IPC and CPU utilization as the efficiency metrics.
\begin{figure}[ht]
\centering
\includegraphics[height=4.5cm]{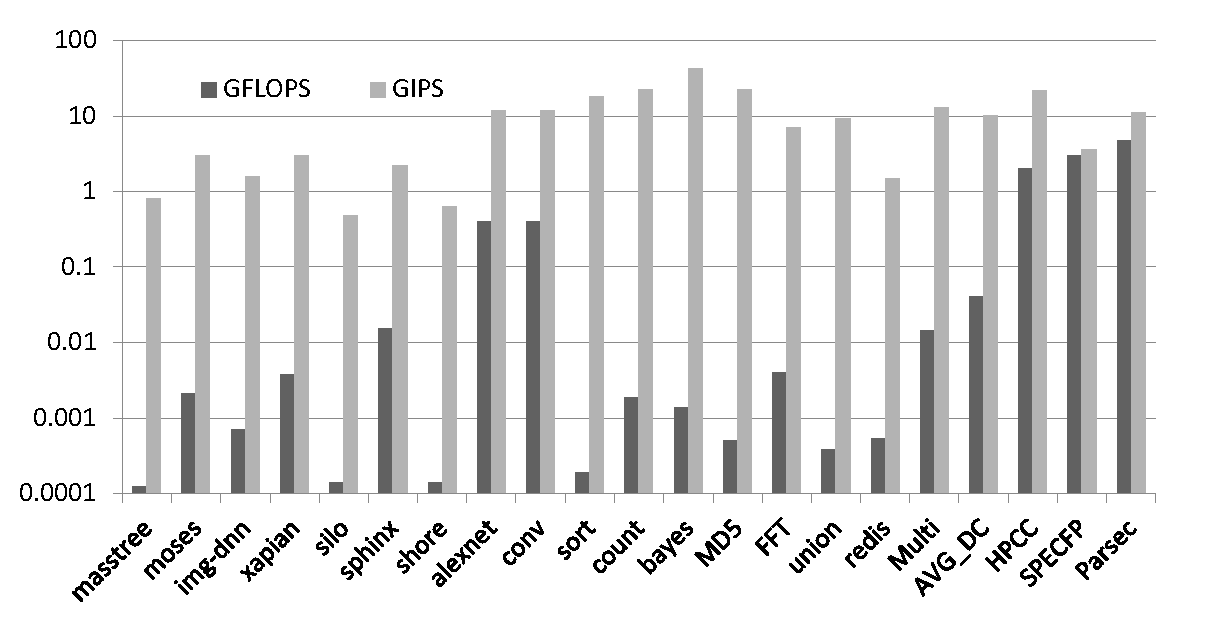}
\caption{GIPS and FLOPS of Workloads.}
\label{mipsfig}
\end{figure}
As shown in the Fig. ~\ref{mipsfig} (please note that the Y axis in the figure is in logarithmic coordinates), the average GFLOPS of DC workloads is two magnitude orders lower than that of traditional benchmarks, while the GIPS of DC workloads is in the same magnitude order as the traditional benchmarks. And \textbf{the average FLOPS efficiency is only 0.1\% for DC workloads}. Furthermore, the average IPC of DC workloads is 1.1 and that of traditional benchmarks is 1.4, the average CPU utilization of DC workloads is 70\% and that of traditional benchmarks is 80\%. These metrics imply that DC workloads can utilize the system resource as efficiently as traditional benchmarks. The poor FLOPS efficiency does not lie in the lower execution efficiency. In fact, the floating point operation intensity of DC workloads (0.05 on average) is much lower, which leads to the low FLOPS efficiency.

\begin{figure}[ht]
\centering
\includegraphics[height=4.0cm]{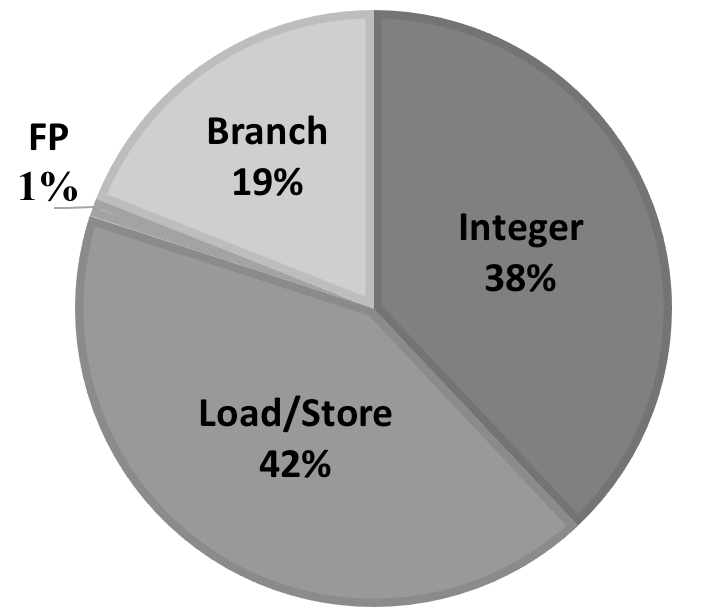}
\caption{Instructions Mixture of DC Workloads. }
\label{piefig}
\end{figure}

In order to analyze the execution characteristics of DC workloads, we choose the instruction mixture to perform the further analysis. Fig. ~\ref{piefig} shows the retired instructions breakdown, and we have three observations as follows. First, the load/store instructions of DC workloads take 42\% of total instructions.
Furthermore, \textbf{the ratio of data movement related instructions is 60\%}, which include the load, store, array addressing instructions (we obtaining the array addressing instructions through analyzing the integer and floating point instructions). \textbf{So, DC workloads are data movement dominated   workloads}. Second, the integer/FP instructions of DC workloads take 39\% of total instructions. Furthermore, \textbf{for DC workloads, the ratio of integer to floating point instructions is 38}, while the ratios for HPCC, Parsec and SPECFP are 0.3, 0.4, and 0.02, respectively. That is the main reason why FLOPS does not work in DC computing. Third, \textbf{DC workloads have more branch (comparing) instructions}, with the ratio of 19\%, while the ratios of HPCC, Parsec and SPECFP are 16\%, 11\%, and 9\%, respectively. So, \textbf{DC workloads are data movement dominated workloads, which have more integer and branch operations}.

\subsubsection{What should be included in our new metric for the DC computing} \label{IMA_DC}
\begin{figure}[ht]
\centering
\includegraphics[height=4.0cm]{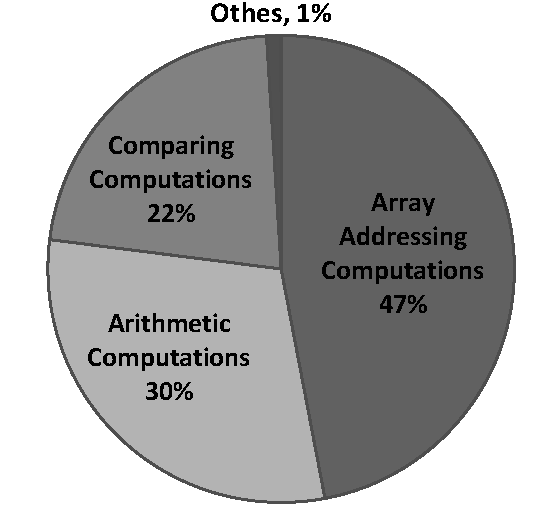}
\caption{Floating point and Integer Operations Breakdown.}
\label{InsBreakfig}
\end{figure}

As the computation-centric metric, we abstract basic computation operations from integer and floating point instruction analysis of DC workloads. First, from the instruction mixture analysis, we know that the ratio of integer to floating point instructions is 38. So, we must consider both integer and floating point operations. Second, as floating point and integer operations are more diverse and all of them are different, we will not consider all of floating point and integer operations. Following the rule to choose a representative minimum operations subset of DC workloads. We analyze the floating point and integer operations of microbenchmarks of DCMIX through inserting the analysis code into the source code. Fig. ~\ref{InsBreakfig} shows the floating point and integer operations breakdown, we find that the 47\% of total floating point and integer operations belong to \textbf{array addressing computations} (data movement related computations operations), 30\% of those belong to \textbf{the arithmetic computations}, 22\% of those belong to \textbf{the comparing computations} (conditional comparing related computation), and others take 1\%. So, we choose data movement computations, arithmetic computations and comparing computations as the minimum operations subset of DC workloads.
\section{BOPS}
BOPS (Basic OPerations per Second) is the average number of BOPs (Basic OPerations) for a specific workload completed per second.
In this section, we present the definition of BOPs and how to measure BOPS with or without the available source code.
\subsection{BOPs Definition}
We summarize basic operations of DC from three classes, which are Data Movement, Arithmetic Computation and Comparing.
\subsubsection{Data Movement} For the FLOPS metric, it is designed for numerical calculation, especially for high floating point operation intensity algorithm, such as the floating point operation intensity ($\textstyle{OI}$) of HPL is O(N), and the data movement can be ignored (one orders of magnitude lower than the floating point operations). On the other hand, in order to process the massive data in time, the complexity of DC workloads are always low, and the operation intensity (the total number of floating point and integer operations divided by the total byte number of memory access) of DC workloads is O(1). So, the data movement can not be ignored. We choose the array addressing computation operations corresponding to data movement-related operations. So, the first class in BOPs is \textbf{array addressing operations}, such as loading or storing array values $\textstyle{P[i]}$.
\subsubsection{Arithmetic Computation} The arithmetic operations is the key operations for the workload's algorithm implementations. We take the basic arithmetic computation operations into BOPs. So, the second class is \textbf{the arithmetic operations}, such as $\textstyle{X+Y}$.
\subsubsection{Comparing} DC workloads have more comparing operations. So we take conditional comparing related computation operations into the BOPs, the third class is \textbf{the comparing operations}, such as $\textstyle{X\textless Y}$.

The detailed operations of BOPs are shown in Table~\ref{Call}. Each operation in Table~\ref{Call} is counted as 1 except for N-dimensional array addressing. Note that all operations are normalized to 64-bit operation. For arithmetic operations, the number of BOPs is counted as the number of corresponding arithmetic operations. For array addressing operations, we take the one-dimensional array $\textstyle{P[i]}$ as the example. Loading the value of $\textstyle{P[i]}$ indicates the addition of an $\textstyle{i}$ offset to the address location of $\textstyle{P}$, so the number of BOPs increases by one. And, it can also be applied to the calculation of the multi-dimensional array. For comparing operations, we transform them to subtraction operations. We take $\textstyle{X \textless Y}$ as an example and transform it to $\textstyle{X-Y \textless 0}$, so the number of BOPs increases by one.
\begin{table}[ht]
\center
\begin{footnotesize}
\caption{Normalization Operations of BOPs.}
\label{Call}
\begin{tabular}{|l|l|}
\hline
   Operations     & Normalized value            \\
  \hline
  Add       & 1            \\
    \hline
 Subtract       & 1            \\
    \hline
 Multiply       & 1            \\
    \hline
 Divide       & 1            \\
    \hline
 Bitwise operation      & 1            \\
    \hline
 Logic operation & 1            \\
    \hline
 Compare operation      & 1            \\
    \hline
 One-dimensional array addressing      & 1            \\
    \hline
 N-dimensional array addressing      & 1*N            \\
    \hline
  \end{tabular}
\end{footnotesize}
\end{table}

Through the definition of BOPs, we can see that in the comparison with FLOPS, BOPS concerns not only the floating-point operations, but also the integer operations. On the other hand, like FLOPs, BOPs normalize all operations into 64-bit operations, and each operation is counted as 1. The delays of different operations are not considered in the normalized calculation of BOPs, because the delays can be extremely different in different micro-architecture platforms. For example, the delay of the division in Intel Xeon E5645 processor is about 7-12 cycles, while in Intel Atom D510 processor, the delay can reach up to 38 cycles~\cite{delayD510}. Hence, the consideration of delays in the normalization calculations will lead to architecture-related issue.
\subsection{How to Measure BOPs}
\subsubsection{Source-code level measurement}
We can calculate BOPs from the source code of a workload, and this method needs some manual work (inserting the counting code). However, it is independent with the underlying system implementation, so it is fair to evaluate and compare different system and architecture implementations. As the following example shows, BOPs is not calculated in Lines 1 and 2, because they are variable declarations. Line 3 consists of a loop command and two integer operations, and the number of corresponding BOPs is (1+1) * 100 = 200 for the integer operations, while the loop command is not calculated; Line 5 consists of the array addressing operations and addition operations: the array addressing operations are counted as 100 * 1, and the addition operations are counted as 100*1, so the sum of BOPs in the example program is: 200 + 200 = 400.
\begin{lstlisting}
1 long newClusterSize[100];
2 long j;
3 for (j=0; j<100;j++)
4 {
5  newClusterSize[j]=j+1;
6 }
\end{lstlisting}

To measure BOPs in the source code level, we need to insert the counting code and the debug flag. To count BOPs,
we will turn on the debug flag, and for the performance evaluation, we will turn off the debug flag.

Another thing we need to take into account is the system built-in library functions. For the calculation of the system-level functions, such as Strcmp() function, we implement user-level functions manually, and then count the number of BOPs through inserting the counting code.

\begin{figure}[ht]
\centering
\includegraphics[height=4.5cm]{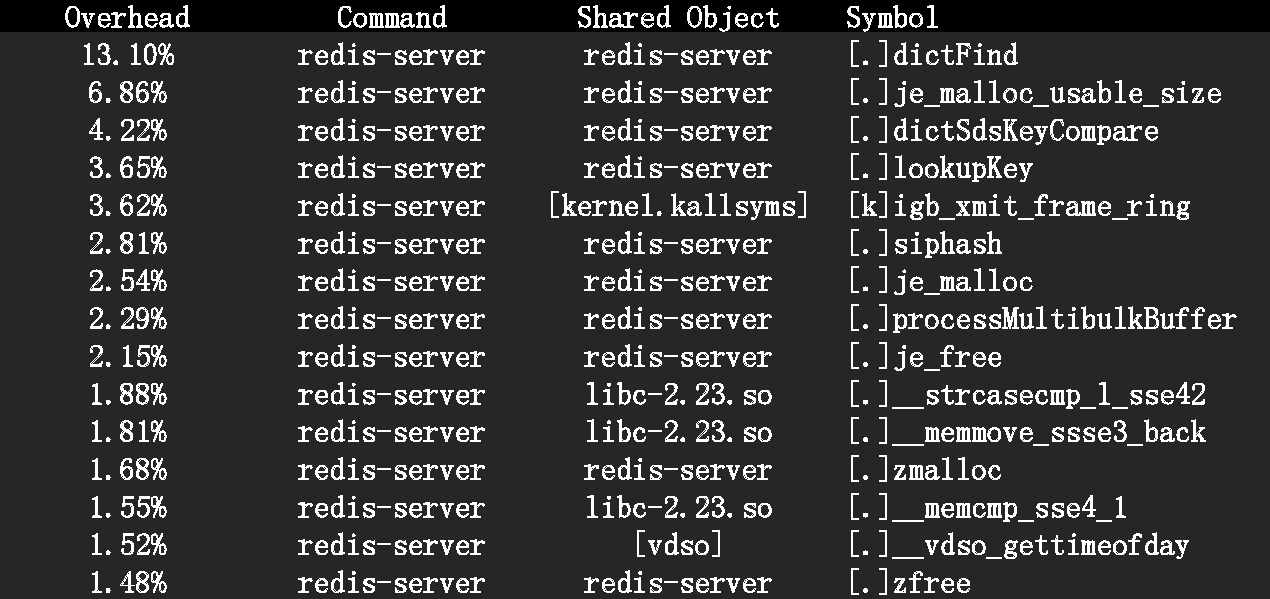}
\caption{Hotspot Functions of the Redis Workload.}
\label{RedisBops1fig}
\end{figure}

For the microbenchmark workloads, we can insert the counting codes easily through analyzing the source code. For the component
benchmarks or real applications, first, we profile the execution time of the real DC workload and find out
the Top N hotspot functions. Second, we analyze these hotspot functions and insert the counting code into these functions.
Then, we can count BOPs for the real DC workload. For example, as shown in the Fig.~\ref{RedisBops1fig}, there are 20 functions
which occupy 69\% execution time of the Redis workload. We insert the counting codes to these hotspot functions, then we can get BOPs through running the Redis workload.
\subsubsection{Instruction level measurement under X86\_64 architecture}
The source-code measurement need to analyze the source code, which costs a lot especially for complex system stacks (e.g., Hadoop system stacks). Instruction level measurement can avoid this high analysis cost and the restriction of needing the source code, but it is architecture-dependent. We propose an instruction-level approach to measure BOPs, which uses the hardware performance counter to obtain BOPs. Since different types of processors have different performance counter events, for convenience, we introduce an approximate but simple instruction level measurement method under X86\_64 architecture. That is, we can obtain the
number of related instructions through the hardware performance counters. And BOPs can be calculated according to the following equation (please note that this equation is for Intel E5645, which equipped with 128-bit SSE FPUs and ALUs).
\begin{sequation}
BOPs=Integer\_All+FP\_All
\end{sequation}
\begin{sequation}
Integer\_All=Integer\_Ins+2*SSE\_Integer
\end{sequation}
\begin{sequation}
FP\_All=FP\_Ins+SSE\_Scalar+2*SSE\_Packed
\end{sequation}
Please note that our instruction level measurement method includes all of floating point and integer instructions under X86\_64
architecture, which does not exactly conform to the BOPS definition. So, it is a approximate measurement method, and does not suit for the performance evaluation among different micro-architectures (such as CISC Vs. RISC). However, on the same Intel X86\_64 platforms, the deviation between the instructions level measurement and the source code level measurement is no more than 0.08, through our experiments on Intel Xeon E5645.
\subsection{How to Measure the system with BOPS}
\subsubsection{The Peak BOPS of the System}
BOPS is the average number of BOPs for a specific workload completed per second. The peak BOPS can be calculated by the micro-architecture with the following equation.
\begin{sequation}
BOPS_{Peak}=Num_{CPU}*Num_{Core}*Frequency*Num_{BOPs Per Cycle}
\end{sequation}
For our Intel Xeon E5645 experimental platform, the CPU number is 1, the core number is 6, the frequency of core is 2.4 GHZ, BOPs per cycle is 6 (The E5645 equips two 128-bit SSE FPUs and three 128-bit SSE ALUs, and according to the execution port design, it can execute three 128-bit operations per cycle). So $\textstyle{BOPS_{Peak}=1*6*2.4G*6=86.4 GBOPS}$.
\subsubsection{The BOPS Measuring Tool}
We provide a BOPS measuring tool to measure the performance of DC systems. At present we choose Sort in the DCMIX as the first BOPS measuring tool. To deal with the diversity of DC workloads, we will develop a series of representative workloads as the BOPS measuring tools. We choose Sort as the first BOPS measuring tool for it is the most widely used workload in the DC~\cite{sort}. And the Sort workload realizes the sorting of an integer array of a specific scale, the sorting algorithm uses quick sort algorithm and the merge algorithm. The program is implemented by C++ and MPI.
The scale of the Sort workload is 10E8 records, and BOPs of that is 529E9. The details of BOPs can be found in the Table~\ref{BOPs1102}, Please note that BOPs value will change as the data scale changes.
\begin{table}[H]
\center
\begin{footnotesize}
\caption{BOPs of the Sort Measuring Tool}
\begin{tabular}{|l|l|}
  \hline
       Operations   &  Counters       \\
  \hline
       Arithmetic operations &   106E9         \\
  \hline
       Comparing operations & 36E9       \\
  \hline
       Array addressing operations & 387E9     \\
  \hline
       Total & 529E9     \\
  \hline
\end{tabular}
\label{BOPs1102}
\end{footnotesize}
\end{table}
\subsubsection{Measuring the System with BOPS}
The measuring tool can be used to measure the real performance of the workload on the specific system. Furthermore, the BOPS efficiency can be calculated by the following equation.
\begin{sequation}
BOPS_{Efficiency}=BOPS_{Real}/BOPS_{Peak}
\end{sequation}
For example, Sort has 529E9 BOPs. We run Sort on the Xeon E5645 platform and the execution time is 18.7 seconds. $\textstyle{BOPS_{Real}=529E9/18.7=28 GBOPS}$. For the Xeon E5645 platform, the peak BOPS is 86.4 GBOPS, the real performance of Sort is 28 GBOPS, so the efficiency is 32\%.
\subsubsection{The Upper Bound Performance Model}
We modify the Roofline model through changing the metric from FLOPS to BOPS, we call the BOPS based upper bound model as DC-Roofline.
\begin{sequation}
BOPS_{AttainedPeak}=min(OI_{BOPS}*MemBand_{Peak},BOPS_{Peak})
\end{sequation}
$\textstyle{BOPS_{Peak}}$ and $\textstyle{MemBand_{Peak}}$ are the peak performance of the platform, and the operation
intensity ($\textstyle{OI_{BOPS}}$) is the total number of BOPs divided by the total byte number of memory access.
For example, the OI of the sort benchmark is 3.0, the peak memory bandwidth is 13.8 GB/s, the peak BOPS is 86.4 GBOPS. So, the attained peak BOPS of the Sort is 41.4 GBOPS and the attained BOPS efficiency is 68\%.
\begin{sequation}
BOPS_{AttainedEfficiency}=BOPS_{Real}/BOPS_{AttainedPeak}
\end{sequation}
\subsubsection{Adding Ceilings for DC-Roofline}
\begin{figure}[ht]
\centering
\includegraphics[height=7cm]{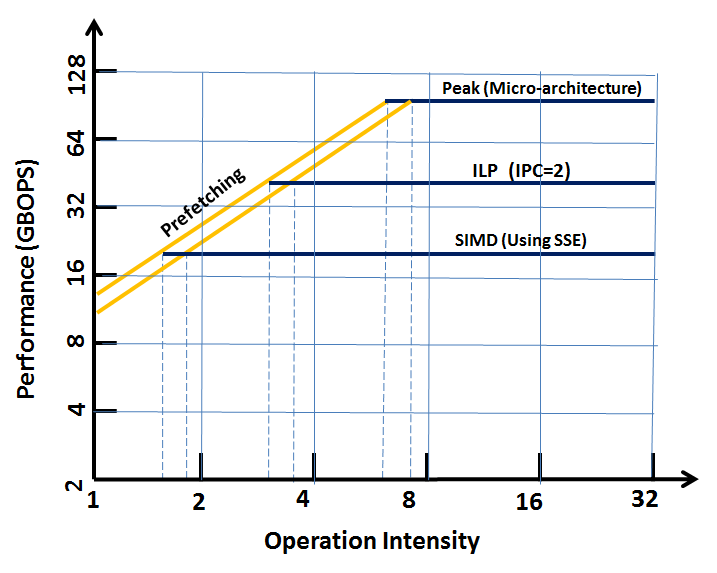}
\caption{Visualized the DC-Roofline model on the Intel E5645 Platform.}
\label{fig:Ceilings}
\end{figure}

We add three ceilings --ILP, SIMD, and Prefetching-- to specify the performance upper bound for the specific tuning settings. Among them, ILP and SIMD reflect computation limitations and Prefetching reflects memory access limitations. We evaluate our experiments on Intel Xeon E5645 platform. We use the Stream benchmark as the measurement tool for the Prefetching ceiling. We improve the memory bandwidth through opening the pre-fetching switch option in the system BIOS, and then the peak memory bandwidth increases from 13.2GB/s to 13.8GB/s. Then, we add two calculation ceilings. The first one is SIMD and the second one is ILP. SIMD is the common method for the HPC performance improvement, which performs the same operation on multiple data simultaneously. Modern processors have 128-bit wide SIMD instructions at least (i.e., SSE, AVX, etc.). In the next sub-section, we will show that SIMD also suits for the DC workload. ILP efficiency can be described by the IPC efficiency (the peak IPC of E5645 is 4). We add the ILP ceiling with IPC no more than 2 (according to our experiments, the IPC of all of workloads is no more than 2), and add the SIMD ceiling with the SIMD upper bound performance. We use the following equation to estimate the ILP and SIMD ceilings:
\begin{sequation}
BOPS_{Ceiling}=BOPS_{Peak}*ILP_{Efficiency}*SIMD_{Scale}
\end{sequation}
where $ILP_{Efficiency}$ is the IPC efficiency of the workload, $SIMD_{Scale}$ is the scale of SIMD (for E5645, the value is 1 under SIMD, the value is 0.5 under SISD). Therefore, ILP (Instruction-level parallelism) ceiling is 43.2 GBOPS when the IPC number is 2. Based on the ILP ceiling, the SIMD ceiling is 21.6 GBOPS when not using SIMD. At last, the attained performance bound of a given workload under ceilings is described as follows.
\begin{sequation}
BOPS_{AttainedC}=Min(BOPS_{Ceiling}, MemBand_{Ceiling}*OI_{BOPS})
\end{sequation}
The visualized DC-Roofline model on the Intel Xeon E5645 platform can be seen in Fig.~\ref{fig:Ceilings}. The diagonal line represents the bandwidth, and the
roof line shows the peak BOPS performance. The confluence of the diagonal and roof lines at the ridge point (where the diagonal and horizontal roofs meet) allows us to evaluate the performance of the system. Similar to the original Roofline model, the ceilings -- which imply the performance upper bounds for the specific tuning settings -- can be added to the DC-Roofline model. There are three ceilings (ILP, SIMD, and Prefetching) in the figure.

\section{Evaluations}
\subsection{Experimental Platforms and workloads}
We choose DCMIX as DC workloads, and choose three typical HPC microbenchmarks (HPL, Graph500, and Stream) as the experimental workloads too. Three systems equipped with three typical Intel processors are chosen as the experimental platforms, which are Intel Xeon E5310, Intel Xeon E5645 and Intel Atom D510. The detailed settings of platforms are shown in Table~\ref{hwconfigeration1}.
\subsection{The BOPS Metrics for DC Systems Evaluations}
\begin{figure}[ht]
\centering
\includegraphics[height=7.0cm]{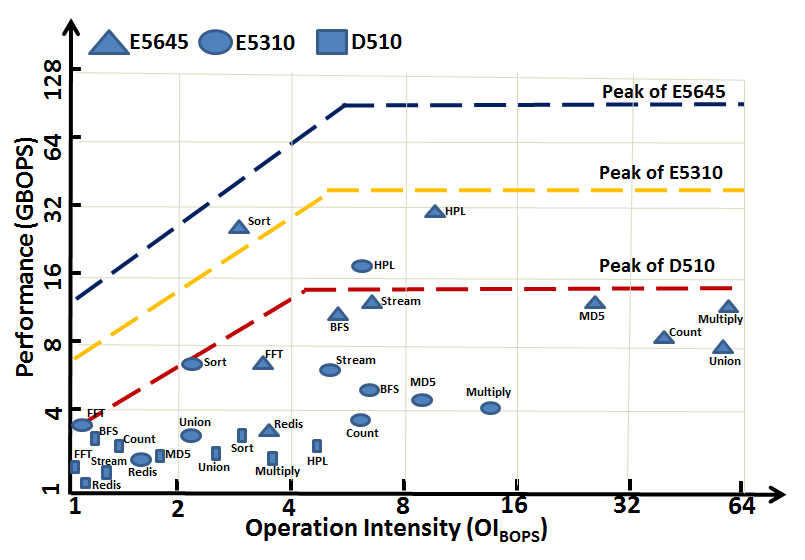}
\caption{Evaluation of Three Intel Processors Platforms with BOPS.}
\label{Bops1fig}
\end{figure}

Fig.~\ref{Bops1fig} is the visualized BOPS based upper bound performance model (Equation 9). There are six DCMIX microbenchmarks, one typical component benchmarks (the Redis workload) and three typical HPC microbenchmarks in the Figure. And three experimental platforms are also in the figure. We see that all of performance metrics are unified to BOPS metric, which include the peak performance of the system (such as the 'Peak of E5645' is the peak performance of the E5645 platform), and the performance of the workload (such as performances of the Sort workload under different platforms). So, we can do the following evaluations. First, analyzing the performance gaps of different systems. Second, performing the apple-to-apple comparison for DC systems. Third, analyzing the upper bound performance of DC systems.

\subsubsection{The Performance Gaps across Different DC Platforms}
Reflecting the performance gaps of different DC systems is the first requirements for BOPS. From Fig.~\ref{Bops1fig}, we see that:

First, for the performance gaps between E5310 and E5645, the peak BOPS performance gap is 2.3X (38.4 GBOPS v.s. 86.4 GBOPS), the gap of the average wall clock time is 2.1X. The bias is only 10\%.

Second, for the performance gaps between D510 and E5645, the peak BOPS performance gap is 6.7X (12.8 GBOPS v.s. 86.4 GBOPS), the gap of the average wall clock time is 7.4X. The bias is only 9\%.

Third, for the performance gaps between D510 and E5310, the peak BOPS performance gap is 3X, the gap of the average wall clock time is 3.4X. The bias is only 10\%.

So, the bias between the peak BOPS performance gap and the average wall clock time gap is no more than 10\%.
\subsubsection{The Apple-to-Apple Comparison of DC systems}
We take the Redis workload (the typical online service workload) and the Sort workload (the typical data analytic workload) as the example to illustrate the apple-to-apple comparison. On the E5645 platform, the Redis workload is 2.9 GBOPS, the performance efficiency of theory peak is 20\% and that of the theory upper bound is 34\% (Redis is a single-threaded server and we deploy it on the single specific CPU core). The Sort workload is 28 GBOPS, the efficiency of the theory peak is 32\% and that of the theory upper bound is 68\% (Sort is a multi-threaded workloads). We see that the Sort workload is more efficiency on the E5645 platform, and we can also do the optimizations base on the upper bound performance model (more details are in the next section). On the other hand, the user-perceived metric of Redis is 122,000 Requests/S and that of Sort is 8.3E6/S (sorting 8.3E6 number elements per seconds), we can not get any insight from these user-perceived metrics. So, we can do the apple-to-apple comparisons with BOPS, whatever they are different type workloads (online services v.s. offline data analytics) or different implements (single-threaded Vs. multi-threaded).
\begin{table}[H]
\center
\caption{The Apple-to-Apple Comparison for DC workloads.}
\begin{tabular}{|c|c|c|c|}
\hline
        & Redis & Sort           \\
  \hline
        BOPS &  2.9G   &    28G       \\
  \hline
        BOPS Efficiency&  20\%  &    32\%       \\
  \hline
        BOPS Attained Efficiency & 34\%   & 68\% \\
  \hline
\end{tabular}
\label{RedisSortBOPS}
\end{table}
\subsubsection{The Upper Bound Performance of DC systems}
We use the Sort measuring tool to evaluate the upper bound performance of DC systems. The peak BOPS is obtained by Equation 7.
The real BOPs values are obtained by the source-code level measurement, and BOPS efficiency is obtained by Equation 8. As shown on the Table~\ref{Flops102}, BOPS efficiencies of E5645, E5310 and D510 are 32\%, 20\% and 21\%, respectively. Furthermore, using the BOPS based upper bound performance model (Equation 9 and Equation 10), we get the BOPS attained efficiency of E5645, E5310 and D510 are 68\%, 49\% and 51\%. So we see that the BOPS value is more reasonable to reflect the peak performance and the upper bound performance of real DC systems. \begin{table}[H]
\center
\begin{footnotesize}
\caption{The BOPS Efficiency of DC Platforms.}
\begin{tabular}{|l|l|l|l|}
\hline
         & E5645 & D510 &  E5310         \\
  \hline
       Peak BOPS  & 86.4G  & 12.8G   &  38.4G       \\
  \hline
       Real BOPS &  28G   &  2.7G   &   7.7G         \\
  \hline
       BOPS Efficiency & 32\%   & 21\%   & 20\%         \\
  \hline
       BOPS Attained Efficiency & 68\%   & 49\%   & 51\%         \\
  \hline
\end{tabular}
\label{Flops102}
\end{footnotesize}
\end{table}
\subsection{BOPS for HPC workloads}
We evaluate the traditional HPC benchmarks with BOPS On the E5645 platform. As shown in the Table.~\ref{FlopsBops}, we chose HPL~\cite{dongarra2003Linpack}, Graph500\cite{Ueno2012Highly} and Stream\cite{Mccalpin1995STREAM} as workloads. Under the BOPS metric, the maximum performance gap of different workloads is no more than 3.4X (41 GBOPS v.s. 12 GBOPS). On the other hand, the maximum FLOPS gap is 77.8X (38.9 GFLOPS Vs. 0.05 GFLOPS). Although these benchmarks are all CPU-intensive workloads (their CPU utilizations are close to 100\%), HPL is mainly about floating point addition and multiplication, while BFS and Stream are mainly about data movement related operations. So, there is larger bias among them under the FLOPS metric, but the bias is small under the BOPS metric. This also implies that BOPS is suited for multiple application domains, especially for the data movement dominated workloads.
\begin{table}[H]
\center
\caption{BOPS for Traditional HPC Workloads.}
\begin{tabular}{|c|c|c|c|}
\hline
         & HPL & Graph500 &  Stream          \\
  \hline
       GFLOPS   & 38.9 & 0.05 &  0.8          \\
  \hline
       FLOPS efficiency   &   68\%&  0.04\% &  0.7\%          \\
  \hline
       GBOPS   &   41&  12 &  13        \\
  \hline
       BOPS efficiency   &   47\%&  18\% &  20\%         \\
\hline
\end{tabular}
\label{FlopsBops}
\end{table}
\section{The BOPS's Use Cases}
\subsection{The BOPS based System Evaluation}
We show the user case to evaluate the system with BOPS. We choose five typical workloads to represent five typical application domains. We choose the Union workload (the OLAP workload), the Redis workload (the OLTP workload), the MatrixMultiply workload (the AI workload), the Sort workload (the BigData workload), and the HPL workload (the HPC workload) as experimental workloads. Please note that based on the BOPS metric, all of workloads and application domains can be extended.

\begin{figure}[ht]
\centering
\includegraphics[height=4.0cm]{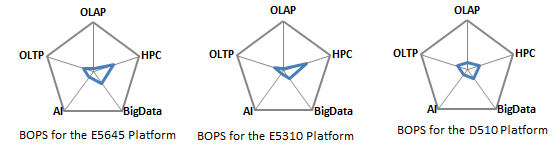}
\caption{The BOPS based System Evaluation.}
\label{Radarfig1}
\end{figure}

We use Kiviat diagrams to show the system evaluation. From Fig.~\ref{Radarfig1}, the average performance efficiencies are 17\%, 20\%, and 23\% for D510, E5310, and E5645 respectively. And the standard deviations are 0.7, 7.1, and 12.8 for D510, E5310, and E5645 respectively. From Fig.~\ref{Radarfig1}, we see that E5310 and E5645 are more suitable for HPC (the HPL workload's BOPS performance is more better than others), while D510 is more balanced (the standard deviations is only 0.7). This is because that both the Xeon E5645 and the Xeon E5310 support out-of-order execution, the four-wide instruction issue, and two 128-bit independent FPUs. On the other hand, D510 is in-order pipeline, two-wide instruction issue, and only one 128-bit multiple FPU. So, E5645 and E5310 are typical Brawny core processors for floating point computations. D510 is Wimpy core processor and more balanced for all application domains.
\subsection{The BOPS based DC Workload Optimizations}
We illustrate how to use the DC-Roofline model to optimize the performance of DC workloads on the E5645 platform. According the upper bound performance model, we adopt four optimization methods. \textbf{[Memory Bandwidth Optimizations]}, we improve the memory bandwidth through opening the pre-fetching switch option. \textbf{[Compiled Optimizations]}, compiled optimization is the basic optimization for the calculation. We improve the calculation performance through adding the compiling option with -O3 (i.e., gcc -O3). \textbf{[OI Optimizations]}, for the same workload, higher OI means the better locality. We modify workload implementations to reduce data movement and increase OI. \textbf{[SIMD Optimizations]}, we apply the SIMD technique to the DC workload, and change programs of the workload from SISD to SIMD through revising SSE. We see that the prefetching optimization and compiled optimization only change the configurations of the system (easy to implement). On the other hand, Operation Intensity optimization and SIMD optimization need to revise the source codes of the workload (hard to implement).

\begin{figure}[ht]
\centering
\includegraphics[height=6.5cm]{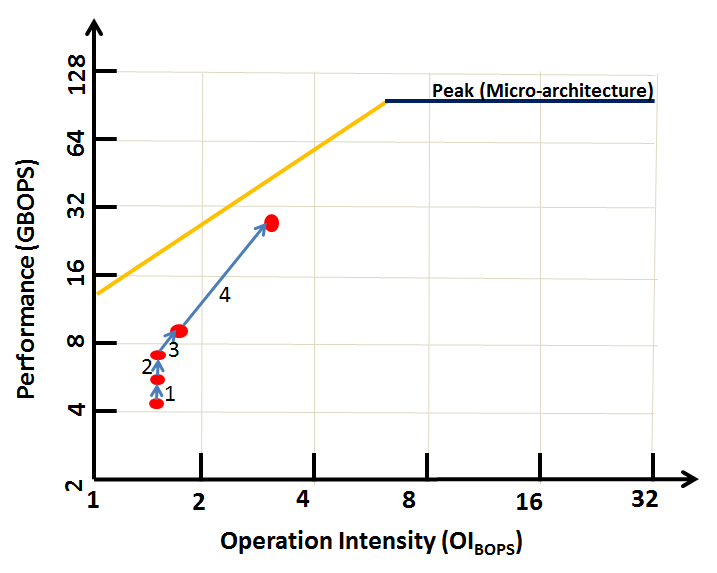}
\caption{Optimizations of the Sort Workload Under the BOPS based Model.}
\label{fig:1112}
\end{figure}

Fig.~\ref{fig:1112} shows the optimization trajectories of the Sort workload. \textbf{In the first step}, we perform Memory Bandwidth Optimization, BOPS increases from 6.4 GBOPS to 6.5 GBOPS. \textbf{In the second step}, we perform Compiled Optimization, improving the performance to 6.8 GBOPS. In the original source codes, data are loading and processing in the disk, and the OI of Sort is 1.4. \textbf{In the third step}, we perform OI Optimization, we revise the source code, which loads and processes all data in the memory, and the OI of Sort increases to 2.2. The performance of Sort increases to 9.5 GBOPS. \textbf{In the forth step},
we apply the SIMD technique to the DC workloads, and change Sort from SISD to SIMD through revising SSE. By using the SEE Sort, the performance is 28 GBOPS, which is 32\% of the peak BOPS. Under the guidance of the DC-Roofline model, the performance improvement of the Sort workload is achieved by 4.4 times.

\begin{figure}[ht]
\centering
\includegraphics[height=7cm]{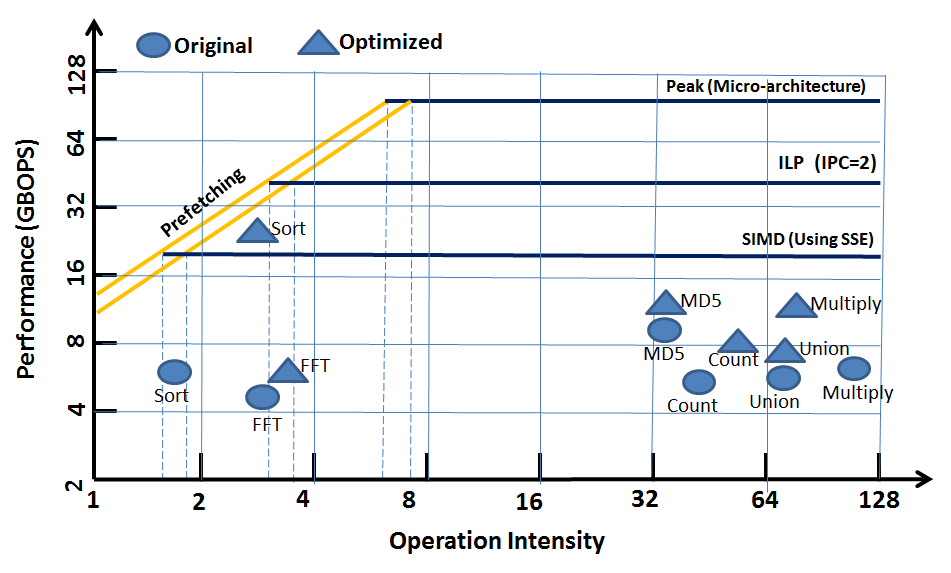}
\caption{DC Workloads' Optimizations on the Intel E5645 Platform.}
\label{E5645RooflineOpt}
\end{figure}

Finally, we take all workloads as a whole to show their performance improvements. For the Sort workload, we execute four optimizations (Memory Bandwidth Optimizations, Compiled Optimizations, OI Optimizations and SIMD Optimizations). For other workloads, we execute two basic optimizations (Memory Bandwidth Optimizations and Compiled Optimizations). As shown in Fig.~\ref{E5645RooflineOpt}, all workloads have achieved performance improvements ranging from 1.1X to 4.4X.
Moreover, we can observe the workload efficiency from the DC-Roofline model. As shown in Fig.~\ref{E5645RooflineOpt}, the workload, which is more closer to the ceiling has higher efficiency. For example, the efficiency of Sort is 65\% under the ILP ceiling, and that of MD5 is 66\% under the SIMD ceiling. The efficiency equation is calculated as follows.
\begin{equation}
BOPS_{CeilingEfficiency}=BOPS_{Real}/BOPS_{AttainedC}
\end{equation}

\begin{figure}[ht]
\centering
\includegraphics[height=7cm]{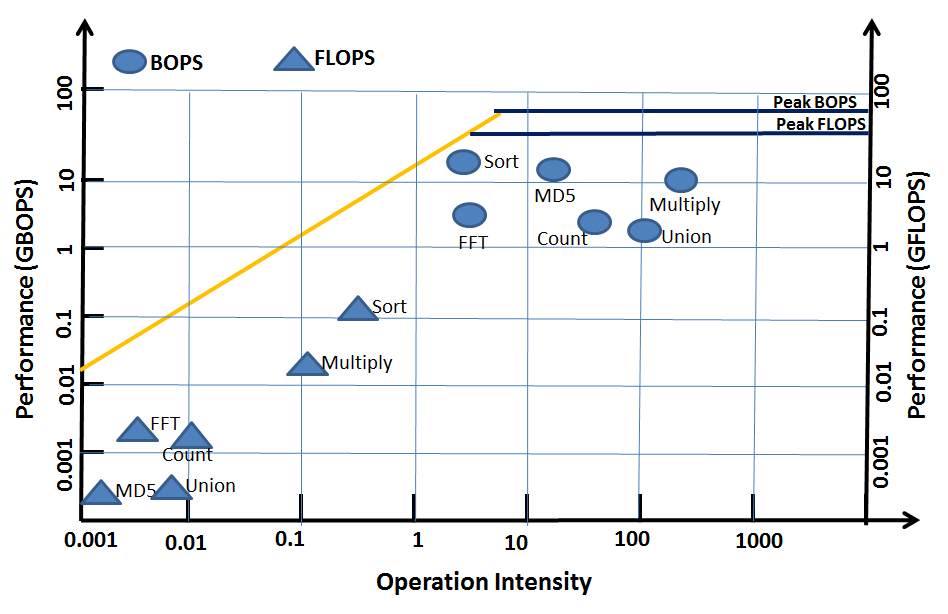}
\caption{The Roofline Model and the DC-Roofline Model on the Intel E5645 Platform. }
\label{DCRfig}
\end{figure}

Fig.~\ref{DCRfig} shows the final results by using Roofline (the left Y axis) and DC-Roofline (the right Y axis), respectively. Please note that the Y axis in the figure is in logarithmic coordinates. From the figure, we can see that if using the Roofline model in terms of FLOPS, the achieved performance is at most up to 0.1\% of the peak FLOPS. For the comparison, the results using the DC-Roofline model is up to 32\% of the peak BOPS. So, the DC-Roofline model is more suited for the upper bound performance model of DC.

\subsection{Optimizing the Real DC Workload Under DC-Roofline Model}
As the real DC workload always has million lines of codes and tens of thousands of functions, it is not easy to use the DC-Roofline model directly (the Roofline or DC-Roofline model is designed for the kernel program optimization). In this section, we propose the optimization methodology for the real DC workload, which is based on the DC-Roofline model. We take the Redis workload as the example to illustrate the optimization methodology, and the experimental results show that the performance improvement of the Redis is 120\% under the guidance of the optimization methodology.
\subsubsection{The Optimization Methodology for the Real DC Workloads}
\begin{figure}[ht]
\centering
\includegraphics[height=3.5cm]{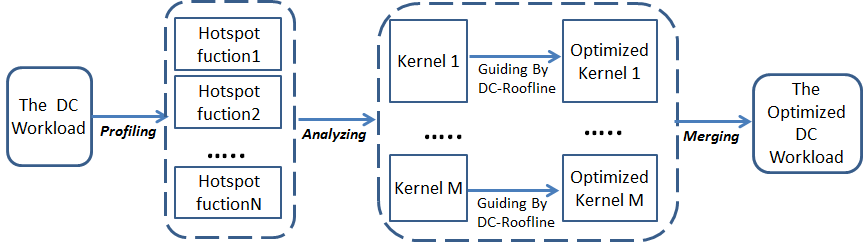}
\caption{The Optimization Methodology for the Real DC workload}
\label{DCMMefig}
\end{figure}

Fig.~\ref{DCMMefig} demonstrates the optimization methodology for real DC workload. First, we profile the execution time of the real DC workload and find out the Top $N$ hotspot functions. Second, we analyze these hotspot functions (merging functions with the same properties) and build the $M$ Kernels ($M$ is less than or equal to $N$). As the independent workload, the Kernel's codes are based on the source code of the real workload and implement a part of functions (specific hotspot functions) of the real workload. Third, we optimize these Kernels through the DC-Roofline model, respectively. Forth, we merge optimization methods of Kernels and optimize the DC workload.
\subsubsection{The Optimization for the Redis Workload}
Redis is a distributed, in-memory key-value database with durability. It supports different kinds of abstract data structures and is widely used in modern internet services. The Redis V4.0.2 has about 200,000 lines of codes and thousands of functions.

\textbf{The Experimental Methodology}

Redis, whose version is 4.0.2, is deployed with stand-alone model. We choose Redis-Benchmark as the workload generator. For the Redis-Benchmark settings, the total request number is 10 millions and 1000 parallel clients are created to simulate the concurrent environment. The query operation of each client is SET operations. We choose the queries per second (QPS) as the user-perceived performance metrics. And the platform is Intel Xeon E5645, which is the same as in Section V.

\textbf{The hotspot Functions of Redis}

There are 20 functions which occupy 69\% execution time. These functions can be classified into three categories. The first category is the dictionary table management, such as dictFind(), dictSdsKeyCompare(), lookupKey(), and siphash(). The second category is the memory management, such as malloc\_usable\_size(), malloc(), free(), zmalloc(), and zfree(). The last category is the encapsulation of system functions. The first two categories take 55\% of total execution time.

\textbf{The Kernels of Redis}

Based on the hotspot functions analysis, we build two Kernels, one is for the memory management, which is called MMK. The other is for the dictionary table management, which is called DTM. Each Kernel is constructed by the corresponding hotspot functions, and can run as an independent workload. Please note that the Kernel and the Redis workload share the same client queries.

\textbf{The Optimizations of DTM}

According to the optimization methods of the DC-Roofline Model (proposed on the Section V), we execute related optimizations.
As the pre-fetching switch option has been opened in the Section V, we execute the following three optimizations: Compiled Optimization, OI Optimization and SIMD Optimization. We perform the Compiled Optimization through adding the compiling option with -O3 (gcc -O3). The optimizations can be shown in Table~\ref{DTM}, we improve
the OI of DTM from 1.5 to 3.5, and BOPS from 0.4 G to 3 G; In the DTM, the rapidly increasing of key-value pairs would trigger the reallocate operation of the dictionary table space, and this operation will bring lots of data movement cost, we propose a method to avoid this operation through pre-allocating a large table space. We call this optimization as NO\_REHASH. Using NO\_REHASH, we improve the OI of DTM from 3.5 to 4, and BOPS from 3 G to 3.1 G; The hash operations are main operations in the DTM, we replace the default SipHash algorithm with the HighwayHash algorithm, which is implemented by the SIMD instruction program. We call this optimization as SIMD\_HASH. Using SIMD\_HASH, we improve the OI of DTM from 4 to 4.7, and BOPS from 3.1 G to 3.7 G.
\begin{table}[H]
\center
\begin{footnotesize}
\caption{Optimizations of DTM}
\label{DTM}
\begin{tabular}{|l|l|l|}
        \hline
        \textbf{Type} & \textbf{OI} & \textbf{GBOPS} \\
        \hline
        Original Version& 1.5 & 0.4  \\
        \hline
        Compiled Optimization & 3.5 & 3  \\
        \hline
        OI Optimization& 4 & 3.2  \\
        \hline
        SIMD Optimization & 4.7 & 3.7  \\
        \hline
\end{tabular}
\end{footnotesize}
\end{table}

\textbf{The Optimizations of MMK}

According to the optimization methods of the DC-Roofline Model, we do two optimizations: Compiled Optimization and OI Optimization. We do the Compiled Optimization through adding the compiling option with -O3 (gcc -O3). The optimization can be shown in Table~\ref{DTM}, we improve the OI of MMK from 3.1 to 3.2, and BOPS from 2.2 G to 2.4 G; To reduce data movement cost, We replace the default malloc algorithm with the Jemalloc algorithm in the MMK. Jemalloc is a general purpose malloc implementation that emphasizes fragmentation avoidance and scalable concurrency support. We call this optimization as JE\_MALLOC. Using JE\_MALLOC, we improve the OI of MMK from 3.2 to 90, and BOPS from 2.4 G to 2.7 G.
\begin{table}[H]
\center
\begin{footnotesize}
\caption{Optimizations of MMK}
\label{MMK}
\begin{tabular}{|l|l|l|}
        \hline
        \textbf{Type} & \textbf{OI} & \textbf{GBOPS} \\
        \hline
        Original Version& 3.1 & 2.2  \\
        \hline
        Compiled Optimization & 3.2 & 2.4  \\
        \hline
        OI Optimization& 90 & 2.7  \\
        \hline
\end{tabular}
\end{footnotesize}
\end{table}

\textbf{The Optimizations of Redis}

\begin{figure}[ht]
\centering
\includegraphics[height=7cm]{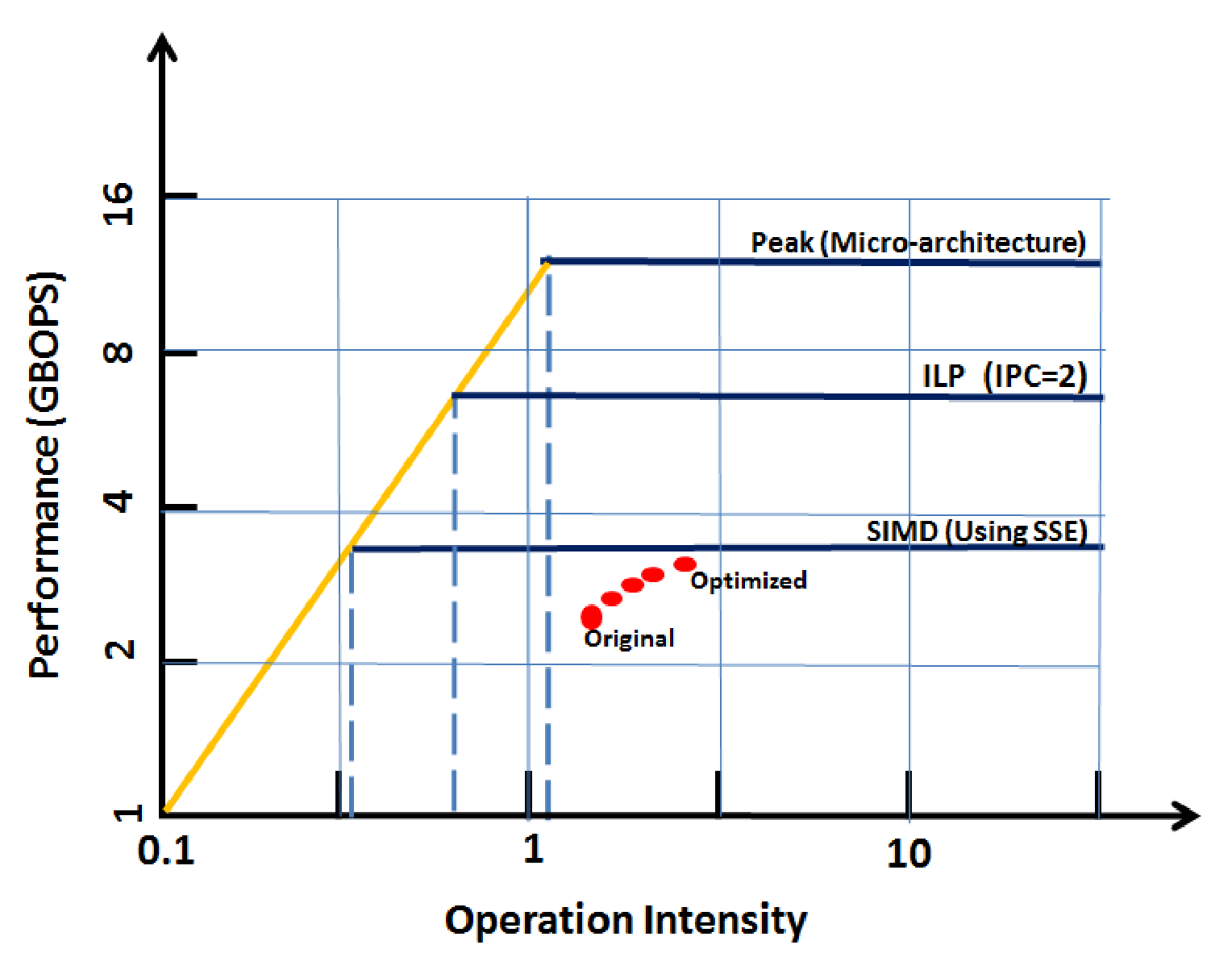}
\caption{Optimization Trajectories of the Redis Workload on the Intel E5645 Platform.}
\label{RedisOpfig}
\end{figure}

Merging the above optimizations of DTM and MMK, we do the optimizations for the Redis workload. Fig.~\ref{RedisOpfig} shows the optimization trajectories, we execute the Compiled Optimization (adding the compiling option with -O3), OI Optimization (NO\_REHASH and JE\_MALLOC), and SIMD Optimization (SIMD\_HASH) one by one. Please note that the peak performance of the system is 14.4 GBOPS in the Fig.~\ref{RedisOpfig}. This is because that the Redis is a single-threaded serve and we deploy it on the single specific CPU core. As shown in Fig.~\ref{RedisOpfig}, the OI of the Redis workload is improved from 2.9 to 3.8, and BOPS from 2.8 G to 3.4 G. And, the QPS of the Redis is improved from 122,000 requests/s to 146,000 requests/s.
\section{Conclusion}
For the system and architecture community, performing the apple-to-apple comparison and obtaining the upper bound performance of the specific system are very important for the system evolution, design and optimization. This paper proposes a new computation-centric metric-—BOPS that measures the DC computing system. The metric is independent
with the underlying systems and hardware implementations, and can be calculated through analyzing the source code. As an effective metric for DC, BOPS can truly reflect not only the performance gaps of different systems, but also the efficiency of DC systems and can be used to perform the apple-to-apple comparison. All of these characteristics are foundations of quantitative analysis for DC systems. And we also illustrate that BOPS can not only guide the optimization of the system (the Sort workload achieved 4.4X performance improvement), but also evaluate the computer system from multiple application domains, which include but are not limited to Big data, AI, HPC, OLTP and OLAP.



\end{document}